\documentclass[prl,twocolumn,showpacs,amsmath,amssymb]{revtex4}
\usepackage[dvips]{graphicx}
\usepackage{bm}

\begin{document}
\title{Full counting-statistics in a single-electron transistor in the presence of nonequilibrium quantum fluctuations of charge}
\author{Yasuhiro Utsumi}
\address{Institut f\"{u}r Theoretische Festk\"{o}perphysik, Universit\"{a}t Karlsruhe, 76128 Karlsruhe, Germany}

\begin{abstract}
Using the Schwinger-Keldysh approach and the drone (Majorana) fermion representation, we evaluate current distribution in a single-electron transistor in a regime where a total tunnel resistance is small. Nonequilibrium quantum fluctuations of charge induce a lifetime broadening of charge-state level, as well as renormalization of system parameters. We find that the lifetime broadening effect may suppress the probability of relatively large current. 
\end{abstract}

\date{\today}
\maketitle

\newcommand{\rd}{d}
\newcommand{\ri}{i}
\newcommand{\mat}[1]{\mbox{\boldmath$#1$}}
\newcommand{\mtau}{\mbox{\boldmath$\tau$}}
\newcommand{\cgf}{{\cal W}}

%
%
Out of equilibrium, 
properties of carrier fluctuations at zero temperature, $T\!=\!0$, 
is beyond the scope of the `fluctuation-dissipation theorem'. 
They have been a highlight of the mesoscopic quantum transport. 
Recently, the \lq full counting statistics' (FCS) 
has pervaded as a powerful concept for 
a description of nonequilibrium current fluctuations~\cite{Qua-Noi-Mes-Phy}. 
The FCS is attractive because it would promote our understanding on 
nonequilibrium strongly-correlated mesoscopic systems. 
However only several attempts have been done for now~\cite{Safi,Kindermann1,Komnik}. 

%
%

A basic example of the strongly-correlated mesoscopic systems
is a \lq single-electron transistor' (SET),
a metallic island coupled to left, right and gate electrodes 
with small total capacitance 
$C_\Sigma\!=\!C_{\rm L}\!+\!C_{\rm R}\!+\!C_{\rm G}$. 
The island can also exchange electrons with left (drain) and right (source) electrodes through tunnel junctions with resistance $R_{\rm L}$ and $R_{\rm R}$. 
However, the charging energy $E_C\!=\!e^2/(2C_\Sigma)$ of a single electron 
can exceed source-drain bias voltage $V$ ($eV\!>\!0$)
and insulate the island [Coulomb blockade (CB)]. 
One can reduce energy for impinging an electron into a charge-neutral island
$\Delta_0 \!=\! E_C(1\!-\!2 Q_{\rm G}/e)$
by gate-induced charge $Q_{\rm G}$. 
Then electrons start to tunnel through the island 
one by one [sequential tunneling (ST)], 
when 
$\Delta_0$ reaches a condition 
$\mu_{\rm R} \!<\! \Delta_0 \!<\! \mu_{\rm L}$,
where
$\mu_{\rm L/R} \!=\! \kappa_{\rm L/R} eV$
[$\kappa_{\rm L/R} \!=\! \pm C_{\rm R/L}(C_{\rm L} \!+\! C_{\rm R})^{-1}$]
is a voltage drop between the  L/R electrode and the island. 

%
%

The physics characteristic for strong electron-correlations 
emerges when a total resistance 
$R_{\rm T} \!=\! R_{\rm L} R_{\rm R}/(R_{\rm L} \!+\! R_{\rm R})$
reduces to the order of the resistance quantum 
$R_{\rm K} \!=\! h/e^2$
and quantum fluctuations of charge are promoted~\cite{Golubev}.
The prime result is renormalization of the 
charging energy and the conductance. 
In the weak tunneling regime where the 
dimensionless conductance 
$\alpha_0 \! \equiv \! R_{\rm K}/(4 \pi^2 R_{\rm T})$ 
is smaller than unity $\alpha_0 \! \ll \! 1$, the scaling analysis predicted 
that the renormalization factor $z_0$ 
logarithmically depends on cutoff energy $\Lambda$, which is
$T$ or $eV$ ($\hbar \!=\! k_{\rm B} \!=\! 1$), 
as 
$z_0 \!=\! 1/ \{ 1 \! + \! 2 \alpha_0 \ln(E_C/\Lambda) \}$~\cite{Matveev}. 

%

For SET-like geometry, the FCS was discussed for 
an open quantum dot coupled to electrodes by single-channel point contacts~\cite{Kindermann1} and for a quantum dot in the Kondo regime~\cite{Komnik}. 
However, for the metallic island, 
the FCS was obtained only in the limit of 
$\alpha_0 \! \rightarrow \! 0$~\cite{Bagrets1}. 
The FCS in the regime where the total resistance is small is still unknown.

%
%

In this paper, we evaluate the FCS of SET
in the weak tunneling regime 
non-perturbatively using the Schwinger-Keldysh approach
and the `drone' (Majorana) fermion representation
~\cite{note1}. 
We will show out of equilibrium, strong quantum charge-fluctuations 
induce a lifetime broadening of charge-state level
and suppress the probability for relatively large current to flow
during a measurement. 
Hereafter, we assume the energy and spin relaxation times 
in each region are fast enough
and electrons obey the Fermi distribution 
$f^+(\omega)\!=\!1/({\rm e}^{\omega/T}\!+\!1)$.

{\it FCS in the weak tunneling regime.} --
In this regime, inverse of the $RC$ time 
$1/\tau
\!=\!
4 \pi \alpha_0 E_C
\!=\! 
1/(R_{\rm T} C_{\Sigma})$ 
is smaller than $E_C$. 
It ensures that charge-state levels are well resolved and the low-energy physics is dominated by charge states with charge neutrality and with one excess electron. 
Then a Hamiltonian is mapped onto  
the \lq multichannel anisotropic Kondo model'~\cite{Matveev},
by introducing a spin-1/2 operator $\hat{\sigma}$ 
acting on the two charge states,
\begin{eqnarray}
\hat{H}
\!=\!
\sum_{\rm r=L,R,I}
\sum_{k n}
\varepsilon_{{\rm r} k} \,
\hat{a}_{{\rm r} k n}^{\dag} \hat{a}_{{\rm r} k n}
\!+\!
\frac{\Delta_0}{2} \hat{\sigma}_z
\nonumber \\
\!+\!
\sum_{\rm r=L,R}
\sum_{k k' n}
(
T_{\rm r} 
\hat{a}_{{\rm I} k n}^{\dag} 
\hat{a}_{{\rm r} k' n}
\hat{\sigma}_{+}
+
{\rm H. c.}),
\label{eqn:H}
\end{eqnarray}
where $\hat{a}_{{\rm r} k n}$ annihilates 
an electron with wave vector $k$ 
in the left or right electrode (r=L,R)
or the island (r=I). 
We assumed the tunneling matrix element 
$T_{\rm r}$ 
is independent of $k$ and the index of channels 
for transverse mode and spin $n$. 
The microscopic parameters and 
the resistance of the junction r 
are related with each other as 
$R_{\rm K}/R_{\rm r} 
\!=\! (2 \pi)^2 N_{\rm ch} |T_{\rm r}|^2 \rho_{\rm I} \rho_{\rm r}$,
where $N_{\rm ch}$ is the number of channels. 
The electron density of states in each region 
$\rho_{\rm r}$ is assumed to be constant.


In the nonequilibrium state, the logarithmic renormalization of 
the system parameters can be treated 
by the `resonant tunneling approximation'~\cite{SchoellerSchoen} 
but only for an {\it average} current. 
A technical difficulty arises because the spin-1/2 operator 
prevents to utilize Wick's theorem. 
A convenient device to avoid the problem might be 
the Majorana fermion representation~\cite{Isawa}, 
$\hat{\sigma}_{+} \! = \! \hat{c}^{\dag} \hat{\phi}$, 
$\hat{\sigma}_{z} \! = \! 2 \hat{c}^{\dag} \hat{c} - 1$
where 
$\hat{\phi} \! = \! \hat{d}^{\dag} + \hat{d}$
is a Majorana fermion and 
$\hat{c}$ and $\hat{d}$ are Dirac fermions. 
The device enables one to utilize the fermionic Schwinger-Keldysh 
approach~\cite{Chou,Kamenev}.

%
%

A central quantity of the technique is  
the generating functional of connected Green function (GF)~\cite{Utsumi}
\begin{equation}
W[\varphi] \! \equiv \! -\ri \ln \int D 
[a_{{\rm r} k n}^*, \!a_{{\rm r} k n}, \!c^*,\!c,\!d^*,\!d ]
\exp \biggl( \ri \int_C \rd t {\cal L}(t) \biggl),
\end{equation}
where ${\cal L}$ is the Lagrangian for $\hat{H}$
and we introduced six Grassmann variables.
The closed time-path $C$ 
advances from $t\!=\!-\infty$ to $\infty$ and 
returns to $-\infty$. 
Then it connects to the imaginary time path
and closes at $t\!=\!-\infty-\ri/T$. 
We introduce an auxiliary source field $\varphi(t)$, 
the phase of the tunneling matrix element 
$T_{\rm r} \rightarrow 
T_{\rm r} {\rm e}^{i \kappa_{\rm r} \varphi(t)}$, 
and assume a part for tunneling is switched on adiabatically. 
Now we can define fields on 
forward and backward branches $\varphi_{+}(t)$ and $\varphi_{-}(t)$, 
but only the center-of-mass coordinate 
$\varphi_c(t) \!\equiv\! \{ \varphi_{+}(t) \!+\! \varphi_{-}(t) \}/2$
has physical meaning and is fixed as $eVt$.

In the scheme of FCS, during measurement time $t_0$, 
the relative coordinate
$\varphi_\Delta(t) \!\equiv\! \varphi_{+}(t) \!-\! \varphi_{-}(t)$ 
is switched on and is fixed as a constant $\lambda$ called 
the `counting field'~\cite{Kamenev}: 
\begin{equation}
\cgf(\lambda) = 
\ri W[\varphi]|_{\varphi_c(t) = eVt,
\varphi_\Delta(t) = \lambda \theta(t_0/2+t) \theta(t_0/2-t)},
\end{equation}
where $\theta(t)$ is the step function. 
Then the generating functional is reduced to the 
cummulant generating function (CGF) for the number of 
transmitted electrons $q$ during a measurement, 
$\cgf(\lambda) 
\!=\! 
\sum_{n=0}^\infty
\langle\!\langle \delta q^n \rangle\!\rangle
(\ri \lambda)^n/n!$. 
The distribution of $q$
(equivalently current $I \!\equiv\!e q/t_0$) 
is given by the inverse Fourier transformation, 
\begin{equation}
P = 
\frac{1}{2 \pi}
\int^{\pi}_{-\pi} 
\!\!\!\!\!\!
\rd \lambda 
\,
{\rm e}^{\cgf(\lambda) -\ri q \lambda}
\approx
{\rm e}^{\cgf(\lambda^*) - \ri (t_0 I/e)  \lambda^*}. 
\label{eqn:invFourier}
\end{equation}
The righthand side is a saddle point solution 
and $\lambda^*$ is obtained by solving 
$I \!=\! -\ri e \partial_\lambda \cgf(\lambda^*)/t_0$~\cite{Bagrets1}. 
The approximation is valid for long measurement time 
since $\cgf$ is proportional to $t_0$ as we will mention later.

We proceed following Ref.~\cite{Utsumi},
where we developed a conserving approximation
for second cummulants, i.e. `noise', 
based on the Schwinger-Keldysh approach
and the Majorana fermion representation. 
We trace out electron degrees of freedom
and 
obtain an effective action for $c$ and $d$ fields
$S^\lambda \! \equiv \!S_{\rm ch} \! + \! S_t^\lambda$
consisting of parts describing the charging energy 
$S_{\rm ch}$ 
and the tunneling $S_t^\lambda$~\cite{Shnirman} as 
\begin{eqnarray}
S_{\rm ch}
\!=\! 
\int _C \! \rd t \{ c(t)^* (\ri \partial_t -\! \Delta_0) c(t) + 
\ri d(t)^* \partial_t d(t) \}, 
\\
S_t^\lambda
=
- \!\!
\int _C \!\!\! \rd t \rd t' \,
c^*(t) \phi(t)
\, \alpha^\lambda(t,t') \,
\phi(t') c(t')
+O(T_{\rm r}^4). 
\label{eqn:tunnelaction}
\end{eqnarray}
The term $O(T_{\rm r}^4)$ describes a phase coherent propagation 
of an electron between the left and right electrodes and can be neglected for 
$N_{\rm ch} \! \gg \! 1$. 
Here
$\alpha^\lambda \!=\! \alpha_{\rm L}^\lambda 
\!+\! \alpha_{\rm R}^\lambda$ 
is a particle-hole (p-h) GF describing 
tunneling of an electron 
from one of the electrodes into the island. 
In the Keldysh space, the p-h GF is expressed 
as a $2 \! \times \! 2$ matrix 
rotated by $\lambda_{\rm r} \!=\! \kappa_{\rm r} \lambda$
along the $x$-axis as
\begin{equation}
\tilde{\alpha}^{\lambda}_{\rm r}(\omega)
\!=\!
U_{\lambda_{\rm r}}^\dagger 
\tilde{\alpha}_{\rm r}(\omega)
U_{\lambda_{\rm r}}, 
\,
\tilde{\alpha}_{\rm r}(\omega)
\!=\!
\biggl(
\!\!
\begin{array}{cc}
0 & \! \alpha_{\rm r}^A(\omega) \\
\alpha_{\rm r}^R(\omega) & \! \alpha_{\rm r}^K(\omega)
\end{array}
\!\!
\biggl),
\label{eqn:p-h}
\end{equation}
where 
$U_{\lambda_{\rm r}}
\!=\!
\exp(- \ri \lambda_{\rm r} \mtau_1/2)$
and
$[\mtau_1]_{ij} \!=\! 1 \! - \! \delta_{ij}$.
Retarded and advanced components are given by 
$
\alpha^R_{\rm r}(\omega) 
\!=\! 
\alpha^A_{\rm r}(\omega)^*
\!=\!
-\ri \pi 
\alpha_0^{\rm r} \, 
(\omega \!-\! \mu_{\rm r}) \,
E_C^2/\{ (\omega \!-\! \mu_{\rm r})^2\!+\!E_C^2 \}$, 
where 
$\alpha_0^{\rm r} \! \equiv \! R_{\rm K}/(4 \pi^2 R_{\rm r})$
and we introduced the Lorentzian cutoff 
to regularize the ultraviolet divergence. 
A Keldysh component 
$\alpha^K_{\rm r}(\omega) \!=\! 
2 \, \alpha^R_{\rm r}(\omega) 
\{ n^+(\omega-\mu_{\rm r}) + n^-(\omega-\mu_{\rm r}) \}$
contains the Bose distribution function
$n^+(\omega) 
\!=\! 1/ ( {\rm e}^{\omega/T} \! - \! 1 )$
and 
$n^-\!=\!1\!+\!n^+$. 

%
%

We observe that only the Dirac fermion $\hat{c}$ may carry 
physical meaning and its free retarded GF 
$g_c^R(\omega) \!=\! 1/(\omega + \ri \eta - \Delta_0)$
describes dynamics of charge-state excitation
($\eta$ is a positive infinitesimal). 
Then the self-energy 
$\Sigma_c^\lambda \!=\! 
\Sigma_{\rm L}^\lambda \!+\! \Sigma_{\rm R}^\lambda$ 
would account for quantum charge-fluctuations caused by tunneling. 
The first order expansion in $\alpha_0$ is 
$\Sigma_{\rm r}^\lambda(t,t') 
\!=\! - \ri g_{\phi}(t',t) \, \alpha_{\rm r}^\lambda(t,t')$,  
where 
$g_\phi$ is the $d$-field GF. 
Keldysh and retarded components of the self-energy for zero 
counting-filed $\tilde{\Sigma}_{\rm r} (\omega)$
are given by
\begin{eqnarray}
\Sigma^K_{\rm r} \! (\omega) \!=\! 2 \alpha^R_{\rm r}(\omega),
\;\;
\Sigma^R_{\rm r}(\omega)
\!=\! \int \!\! \rd \omega' 
\frac{\ri \alpha^K_{\rm r}(\omega')}
{ (2 \pi) (\omega + \ri \eta - \omega') }. 
\end{eqnarray}
%
The integration results in a logarithm as 
$\Sigma_c^R(\omega) \!\approx\! 
\alpha_0 \ln \{ 2 E_C/(eV) \} \, \omega - \ri \Gamma/2$ 
for $|\omega| \! \ll \! eV$
($T\!=0\!$
and a symmetric case, i.e. 
$R_{\rm L} \!=\! R_{\rm R}$ and 
$C_{\rm L} \!=\! C_{\rm R}$). 
The imaginary part describes the lifetime broadening of charge-state level caused by dissipative current:
$\Gamma$ is the sum of tunneling rates 
$\Gamma \! = \!
\Gamma_{\! \rm I L} \!+\! \Gamma_{\! \rm L I} 
\!+\!
\Gamma_{\! \rm I R} \!+\! \Gamma_{\! \rm R I}$, 
where 
$\Gamma_{\! \rm rI/Ir} \!=\!
(\Delta_0 \! - \! \mu_{\rm r}) 
n^\pm(\Delta_0 \! - \! \mu_{\rm r})/(e^2 \! R_{\rm r})$ 
is the tunneling rate into (out of) the island 
through the junction r calculated within Fermi's golden rule.

We can perform a systematic perturbative expansion in $\alpha_0$ 
for the CGF using diagrams~\cite{Utsumi}. 
For example, the first order expansion is given by
$\cgf^{[1]}(\lambda) \!=\! - \! \int_C \! \rd t \rd t'
g_c(t,t') \Sigma^\lambda_c(t',t)$. 
We proceed by projecting the time on $C$ onto the real axis. 
For enough long measurement time, 
we can approximate an integral 
$\delta_{t_0}\!(\omega)
\!\! \equiv \!\!
\int^{t_0/2}_{-t_0/2} \rd t \, {\rm e}^{-\ri \omega t}/(2 \pi)$
and 
its square normalized by $t_0$, 
$2 \pi \delta_{t_0}\!(\omega)^2/t_0$, 
as the $\delta$-function 
$\delta(\omega)$. 
Especially, the latter equation ensures 
that $\cgf$ is proportional $t_0$. 
By performing the Fourier transformation, 
we obtain
\begin{eqnarray}
\cgf^{[1]}(\lambda)
\! \approx \!
- t_0 \! \int \! \rd \omega {\rm Tr}
\{ \tilde{g}_c(\omega) \mtau_{\! 1} 
\tilde{\Sigma}^\lambda_c(\omega) \mtau_{\! 1} \} /(2 \pi)
\nonumber \\
\! = \!
t_0 
\sum\nolimits_{\rm r=L,R}
\{ 
P_- \Gamma_{\! \rm r I} ({\rm e}^{\ri \lambda_{\rm r}}\!-\!1)
\!+\!
P_+ \Gamma_{\! \rm I r} ({\rm e}^{-\ri \lambda_{\rm r}}\!-\!1) \},
\label{eqn:first}
\end{eqnarray}
where we utilized the expression 
of Keldysh component for free $c$-field GF 
$g_c^K(\omega) \!=\! 2 \ri \, {\rm Im} \, 
g_c^R(\omega)(P_{\! -} \! - \! P_{\! +})$. 
It contains initial distribution probabilities of charge state
$P_{\! \pm} \!=\! 1/({\rm e}^{\pm \Delta_0/T}\!+\!1)$.
Equation (\ref{eqn:first}) is the sum of 
CGFs of Poissonian distribution 
describing tunneling of an electron into and out of the island. 
Here, we observe that the naive 
first order expansion Eq.~(\ref{eqn:first}) is insufficient:
First, it includes the initial distribution probabilities,
which should be replaced with stationary distribution probabilities. 
Second, the CGF should depend only on 
$\lambda_{\rm L} \!-\! \lambda_{\rm R} \!=\! \lambda$ 
from the charge conservation~\cite{Belzig}. 
Moreover, the first order expansion is obviously insufficient
to reproduce the logarithmic behavior 
of the scaling analysis. 

Above flaws can be removed by accounting for 
a subclass of diagrams. 
Practically, we simply sum up the geometric series in 
$(\tilde{g}_c \mtau_{\! 1} \tilde{\Sigma}^\lambda_c \mtau_{\! 1})$,
which contains leading logarithms, 
i.e. powers of $\alpha_0 \ln \{ 2 E_C/(eV) \}$: 
\begin{eqnarray}
\cgf(\lambda)
\approx
\frac{t_0}{2 \pi}
\!\! \int \!\! \rd \omega
\, {\rm Tr} \, 
\ln \! \left[
{\tilde{g}_c(\omega)}^{-1}
\!\! -
\mtau_{\! 1}
\tilde{\Sigma}_c^\lambda(\omega)
\mtau_{\! 1}
\right]
\nonumber
\\
=
\frac{t_0}{2 \pi}
\!\! \int \!\! \rd \omega
\ln
\{
1+
T^F\!(\omega) 
f^+\!(\omega \!-\! \mu_{\rm L}) 
f^-\!(\omega \!-\! \mu_{\rm R}) 
({\rm e}^{\ri \lambda}\!-\!1)
\nonumber \\
+
T^F\!(\omega) 
f^+\!(\omega \!-\! \mu_{\rm R}) 
f^-\!(\omega \!-\! \mu_{\rm L})
({\rm e}^{-\ri \lambda}\!-\!1)
\}+O(\eta), \;
\label{eqn:CGF-RTA}
\end{eqnarray}
where $f^- \!=\! 1\!-\!f^+$
and we removed a constant to fulfill 
the normalization condition $\cgf(0)\!=\!0$. 
Now $P_\pm$ is absorbed in $O(\eta)$ and
only $\lambda$ appears in the expression. 
Equation~(\ref{eqn:CGF-RTA}) looks similar to the 
Levitov-Lesovik formula~\cite{Levitov1}. 
But for our case the effective transmission probability 
$T^F \! (\omega)
\! = \! -\alpha_{\rm L}^K \! (\omega) \alpha_{\rm R}^K \! (\omega)/
|\omega - \Delta_0 - \Sigma_c^R(\omega)|^2$
accounts for strong quantum charge-fluctuations. 

In fact, Eq.~(\ref{eqn:CGF-RTA}) is reduction of 
an approximate Keldysh generating functional
Eq.~(25) in Ref.~\cite{Utsumi} to the CGF. 
Therefore, first and second cummulants, 
$\langle I \rangle 
\!=\!
e \langle\!\langle \delta q \rangle\!\rangle/t_0$ 
and 
$S_{\!I\!I} \!=\! 
2 e^2 \langle\!\langle \delta q^2 \rangle\!\rangle/t_0$,
reproduce an average current and a
zero-frequency noise in Ref.~\cite{Utsumi}. 
Especially, for the average current,
our approximation and 
the `resonant tunneling approximation'~\cite{SchoellerSchoen}
are equivalent.

{\it Limiting cases. --}
In the limit of $\alpha_0 \! \rightarrow \! 0$, 
we confirm that 
Eq.~(\ref{eqn:CGF-RTA}) reproduces
the `orthodox' theory~\cite{Bagrets1}:
\begin{eqnarray}
\cgf^{(1)}(\lambda)
\! = \!
t_0 \Gamma(\sqrt{D(\lambda)}-1)/2,
\\
D(\lambda)
\!=\! 
1 +
4 \, \frac{\Gamma_{\! \rm L I} \Gamma_{\! \rm I R}}{\Gamma^2}
({\rm e}^{i \lambda}\!-\!1)
+
4 \, \frac{\Gamma_{\! \rm R I} \Gamma_{\! \rm I L}}{\Gamma^2}
({\rm e}^{-i \lambda}\!-\!1).
\end{eqnarray}
It is noticed that 
though $\cgf^{(1)}$ is proportional to $\alpha_0$, 
it is different from 
the naive first order expansion $\cgf^{[1]}$. 
%
%
The second order expansion in $\alpha_0$ reads, 
\begin{equation}
\cgf^{(2)}(\lambda) \! = \! \cgf^{\rm cot}(\lambda)
\!+\!
\partial_{\Delta_0} \{{\rm Re} \Sigma^R_c(\Delta_0) \cgf^{(1)}(\lambda) \}. 
\label{eqn:CGF-2nd}
\end{equation}
The second term provides the renormalization 
of the system parameters up to first order in $\alpha_0$~\cite{Braggio}. 
The first term is the CGF of a bidirectional Poissonian process
\begin{equation}
\cgf^{\rm cot}(\lambda) \! = \!
t_0 \{ 
\gamma^+ ({\rm e}^{i \lambda}\!-\!1) 
+ 
\gamma^- ({\rm e}^{-i \lambda}\!-\!1) \}, 
\end{equation}
governed by the cotunneling rate, 
$\gamma^\pm \!=\!
2 \pi \,
\alpha_0^{\rm L} \, \alpha_0^{\rm R}
\int \! \rd \omega
(\omega \!-\! \mu_{\rm L}) 
(\omega \!-\! \mu_{\rm R}) 
n^\pm(\omega \! - \! \mu_{\rm L})
n^\mp(\omega \! - \! \mu_{\rm R})
{\rm Re} \{ (\omega + \ri \eta - \Delta_0)^{-2} \}$. 
The first term is relevant in the CB regime and is consistent with 
the FCS theory of quasiparticle tunneling~\cite{Levitov2}. 
Since a tunneling quasiparticle is 
an electron tunneling from 
the left electrode to the right electrode
with a p-h pair left in the island, 
the tunneling rate is not proportional 
to $|T_{\rm L/R}|^2$ 
but $|T_{\rm L} T_{\rm R}|^2$. 

%
%

\begin{figure}[ht]
\includegraphics[width=1.0 \columnwidth]{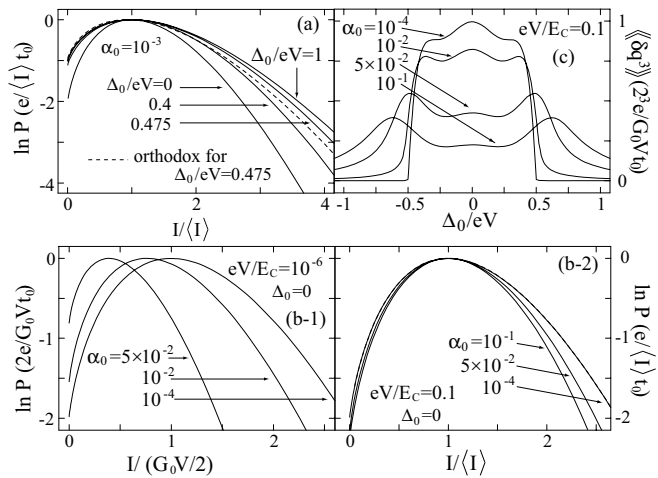}
\caption{
The zero-temperature current distribution 
for $R_{\rm L} \!=\! R_{\rm R}$ 
and 
$C_{\rm L} \!=\! C_{\rm R}$
[(a), (b-1,2)]. 
(a) Solid lines are plots for various excitation energies for 
small conductance.
A dashed line is a result of the `orthodox' theory at $\Delta_0/(eV) \!=\! 0.475$. 
Plots at $\Delta_0 \!=\!0$ for various conductances 
with small (b-1) and 
large (b-2) bias voltages. 
Axes are normalized by 
the average current $\langle I \rangle$ (b-2)
and that of the `orthodox' theory $G_0 V/2$ (b-1). 
(c)
The skewness at $eV\!=\!0.1 E_C$ for various conductances. 
}
\label{fig:distribution}
\end{figure}

We will discuss the saddle point solution 
of the current distribution $P$, Eq.~(\ref{eqn:invFourier}). 
Solid lines in Fig.~\ref{fig:distribution}(a) 
show the logarithm of 
nonequilibrium distribution for the symmetric case at $T\!=\!0$. 
The conductance is small and thus 
$\cgf^{(1)}$ and $\cgf^{\rm cot}$
well approximate CGFs deep in the ST regime 
and in the CB regime, respectively. 
%
%
At $\Delta_0 \!=\!0$, the CGF is 
$\cgf^{(1)}(\lambda) \!\approx\!
2 \, \bar{q} \, ({\rm e}^{\ri \lambda/2}\!-\!1)$, 
where $e \bar{q}/t_0 \!=\! G_0 V/2$
and $1/G_0\!=\!R_{\rm L} \! + \! R_{\rm R}$. 
The factor ${\rm e}^{\ri \lambda/2}$ results in 
a sub-Poissonian value of the Fano factor 
$S_{II}/(2 e \langle I \rangle) \! \approx \! 1/2$
indicating that tunneling processes are correlated. 
The meaning of the `correlated process' can be understood
from an explicit form of distribution 
$P(q) \!=\! 
\sum_{q_{\rm L},q_{\rm R}=0}^\infty
P_{\rm P}(q_{\rm L}) 
P_{\rm P}(q_{\rm R}) 
\, \delta_{q,(q_{\rm L}+q_{\rm R})/2}$
obtained by simply 
performing the inverse Fourier transformation 
Eq.~(\ref{eqn:invFourier}) without the saddle point approximation.
For symmetric case, numbers of 
transmitted electrons through the junctions L and R, 
$q_{\rm L}$ and $q_{\rm R}$, follow 
the same Poissonian distribution 
$P_{\rm P}(q) 
\!=\!
\bar{q}^{\, q} {\rm e}^{-\bar{q}}/q! \,$. 
The Kronecker delta implies that
$q_{\rm L}$ and $q_{\rm R}$ are not independent variables.

As $\Delta_0$ increases and approaches 
a threshold value $\Delta_0 \!=\! eV/2$, 
the left junction becomes dominant 
$\cgf^{(1)}(\lambda)
\! \approx \!
t_0 \Gamma_{\! \rm L I} \, ({\rm e}^{\ri \lambda}\!-\!1)$
and the Poissonian value is approached 
$S_{\!I\!I}/(2 e \langle I \rangle) \!\approx\!1$. 
The Fano factor continues to be unity in the CB regime, 
because the CGF is also that of the Poissonian distribution $\cgf^{\rm cot}$. 
However, since tunneling quasiparticles are different
in two regimes, 
near the threshold, 
there is a regime where neither 
the `orthodox' theory $\cgf^{(1)}$
nor 
the `cotunneling theory' $\cgf^{\rm cot}$
does not work. 
The plot for $\Delta_0/eV\!=\!0.475$ in Fig.~\ref{fig:distribution}(a) 
is such an example. 
The `orthodox' theory [dashed line] cannot fit the solid line. 

{\it Renormalization and lifetime broadening effects. --}
As the conductance increases, the quantum charge-fluctuations are promoted. 
However, when the lifetime broadening is negligible,  
$z_0 \Gamma \! \ll \! eV$ 
[$\Lambda\!=\!\max(|z_0 \Delta_0|,2 \pi T, |eV|/2)$], 
the `orthodox' theory $\cgf^{(1)}$ with 
renormalized parameters $z_0 \alpha_0$ and $z_0 \Delta_0$
is valid.  
Figure~\ref{fig:distribution}(b-1), 
the current distribution at $\Delta_0 \!=\!0$ and
a small bias voltage, indicates the renormalization effect:
Since $z_0$ decreases with increasing $\alpha_0$, 
the mean value, i.e. a peak position, shifts leftwards. 
The above scenario could fail in a regime
$T_{\rm K}\!=\!E_C{\rm e}^{-1/(2 \alpha_0)}/ (2 \pi)
\! \gg \! \Lambda$
where leading logarithms would be not enough. 

%
%

If we replot Fig.~\ref{fig:distribution}(b-1) 
with vertical and horizontal axes normalized
not by $G_0 V/2$ but by $\langle I \rangle$, 
three curves almost overlap (not shown),
i.e. the renormalization factor is normalized away. 
A qualitative change occurs 
when the bias voltage increases and the lifetime broadening effect is enhanced 
[Fig.~\ref{fig:distribution}(b-2)]. 
We observe a shrinking of the distribution with increasing $\alpha_0$. 
The shrinking is compatible with the suppression of 
the Fano factor predicted in Ref.~\cite{Utsumi}.
The FCS analysis gives more information beyond the noise analysis. 
From the current distribution, we can see the probability of 
relatively large current is suppressed. 

Let us discuss the lifetime broadening effect 
for symmetric case more quantitatively. 
For $eV \! \gg \! T_{\rm K}$, 
the real part of $\Sigma^R_c$ can be neglected and 
only the imaginary part describing the lifetime broadening is important, 
$\Sigma^R_c(\omega) \!\approx\!- \ri \pi \alpha_0 eV$. 
Then the CGF at $\Delta_0\!=\!0$ is estimated as
\begin{eqnarray}
\cgf(\lambda) &\approx&
2 \, \bar{q}
\{
({\rm e}^{\ri \lambda/2}\!-\!1)
-2 \alpha_0 ({\rm e}^{\ri \lambda}\!-\!1)
\nonumber \\
&+& 
\pi^2 \alpha_0^2 
({\rm e}^{\ri 3 \lambda/2}\!-\!{\rm e}^{\ri \lambda/2})/2
+O(\alpha_0^3)
\}, 
\label{eqn:CGFexpansion}
\end{eqnarray}
and the `Fano factor' reads
$
\langle\!\langle \delta q^n \rangle\!\rangle/
\langle\!\langle \delta q \rangle\!\rangle
\!=\! 
2^{1-n} 
\{1 \!-\! 4 \alpha_0(2^{n-1} \!-\! 1) 
+ O(\alpha_0^2) \}$.
We can see higher cummulants are suppressed 
as $\alpha_0$ increases.
Figure~\ref{fig:distribution}(c) shows
the skewness $\langle\!\langle \delta q^3 \rangle\!\rangle$
as a function of $\Delta_0$. 
A double-peak structure growing with increasing the conductance 
is obtained. 
Such a characteristic structure might be measurable
for a present-day experiment~\cite{Reulet}. 

Summing up, we evaluated the FCS of SET. 
The strong nonequilibrium quantum charge-fluctuations 
are accounted for by summing up a subclass of diagrams
including leading logarithms. 
Our approximation reproduces 
the `orthodox' theory in the limit of $\alpha_0\!\rightarrow\!0$
and gives the bidirectional Poissonian distribution 
governed by the cotunneling rate in the CB regime. 
Nonequilibrium quantum charge-fluctuations 
induce not only the logarithmic renormalization of 
the conductance and the charging energy 
but also the lifetime broadening of the charge-state level.  
It may suppress the probability of relatively large current and 
higher cummulants of current fluctuations.


We would like to thank 
D. Bagrets,
A. Braggio,
Y. Gefen,
D. Golubev, 
J. K{\"o}nig,
G. Sch{\"o}n 
and 
A. Shnirman
for valuable discussions.
Y.U. was supported by the DFG-Forschungszentrum CFN.


\begin{thebibliography}{19}
\bibitem{note1}
A complete version covering also the strong tunneling regime 
$R_{\rm K} \! \gtrsim \! R_{\rm T}$ will be published elsewhere.
\end{thebibliography}
\end{document}